\documentclass[12pt,preprint]{aastex}

\slugcomment{Draft version.}

\shorttitle{Compton telescope with coded aperture mask}
\shortauthors{Forot et al.}

\begin{document}

\title{Compton telescope with coded aperture mask: \\
    Imaging with the INTEGRAL/IBIS Compton mode }

\author{M. Forot\altaffilmark{1}, P. Laurent\altaffilmark{1}, F. Lebrun\altaffilmark{1} and O. Limousin\altaffilmark{1}}
\altaffiltext{1}{Service d'Astrophysique, CEA Saclay, 91191, GIF sur YVETTE, France}

\email{mforot@cea.fr}

\begin{abstract}
Compton telescopes provide a good sensitivity over a wide field of
view in the difficult energy range running from a few hundred keV
to several MeV. Their angular resolution is, however, poor and
strongly energy dependent. We present a novel experimental design
associating a coded mask and a Compton detection unit to overcome
these pitfalls. It maintains the Compton performance while
improving  the angular resolution by at least an order of
magnitude in the field of view subtended by the mask. This
improvement is obtained only at the expense of the efficiency that
is reduced by a factor of two. In addition, the background
corrections benefit from the coded mask technique, i.e. a
simultaneous measurement of the source and background. This design
is implemented and tested using the IBIS telescope on board the
INTEGRAL satellite to construct images with a $12'$ resolution
over a 29°x29° field of view in the energy range from $200$ keV to
a few MeV. The details of the analysis method and the resulting
telescope performance, particularly in terms of sensitivity, are
presented.

\end{abstract}

\keywords{gamma rays:observation --  instrument:Compton -- telescopes}

\section{Introduction}

The development of Compton telescopes began in the 1970's with
balloon flights\\ \citep{Schoenfelder1973,Herzo1975, Lockwood1979}
and culminated with the flight of COMPTEL \citep{Schoenfelder1993}
on board the Compton Gamma Ray Observatory (CGRO). COMPTEL has
shown for more than 9 years the capabilities of a Compton
telescope to image the sky between 1 and 30 MeV thanks to the
Compton kinematics information \citep{Boggs2000}. The study of
astrophysical sites of nucleosynthesis, as illustrated by the
first $^{26}$Al skymap \citep{Diehl1995}, largely progressed with
the COMPTEL data. On the other hand, COMPTEL barely achieved a 5° deg 
(FWHM) angular resolution at 1 MeV. Future Compton telescopes could
benefit from the very significant detector progress, particularly
in the semiconductor domain, to improve the spectral resolution,
thus the angular resolution \citep{Limousin2003}. The latter is,
however, intrinsically limited by what is referred to as the
electron Doppler broadening which results from the fact that the
scattering electron in a detector is bound. This limits the
angular resolution to about 5 deg at 511 keV in the best case \citep{Zoglauer2003}.

One way to overcome this limitation is to adjoin a coded aperture
(mask) to a Compton telescope. This design, which has never been
used for gamma-ray space telescopes, maintains the advantages of a
Compton telescope (high-energy response, low background, wide
field of view) over most of the wide field of view, but it adds
the coded mask imaging properties (angular resolution, background
subtraction) in the solid angle subtended by the mask. Indeed, in
a coded mask system, source and background are measured
simultaneously and the energy independent angular resolution is
more than one order of magnitude better than in classical Compton
telescopes. With Coded Aperture Compton Telescopes (hereafter
CACT), we can obtain low background images in the 200 keV--10 MeV
energy range, with an angular resolution better than a fraction of
degree (e.g. 10 arcmin). 

Anyway, a CACT has generally a lower
efficiency than a classical coded mask telescope with, for instance,
a thicker detector layer. However, the background in the energy range
 from $200$ keV to a few MeV is dominated by the telescope internal 
emission, which increases with the detector volume. A thicker detector
will suffer thus an higher background. Therefore, the decision
to use only one layer or two layers in coincidence for detecting the photons
through the mask, is a trade-off
between the detector(s) background level and efficiency. 

In another hand,
CACT can be used to get the full energy deposit of an high energy photon, 
even if imaging is done only using the mask projection on one layer. This 
will enable the users to get an improved energy response of a coded mask telescope.

Lastly, CACT are in general difficult to use at higher energy, near 10 MeV and above, as they would require a thick mask, even with tungsten, to stop the photons, which will result in a large vignetting effect for off-axis sources. Also, at those energies, Compton scatterings and pair creation in the mask will affect the system imaging properties, and may degrade its angular resolution which therefore becomes slightly energy dependant.

In this paper, we  present the general
principle of CACT, their application to the INTEGRAL mission, the
difficulties inherent to the use of CACT, the analysis method of
the IBIS Compton Mode and its resulting performance.
\clearpage
\begin{figure}[!h]
\epsscale{1}
\plotone{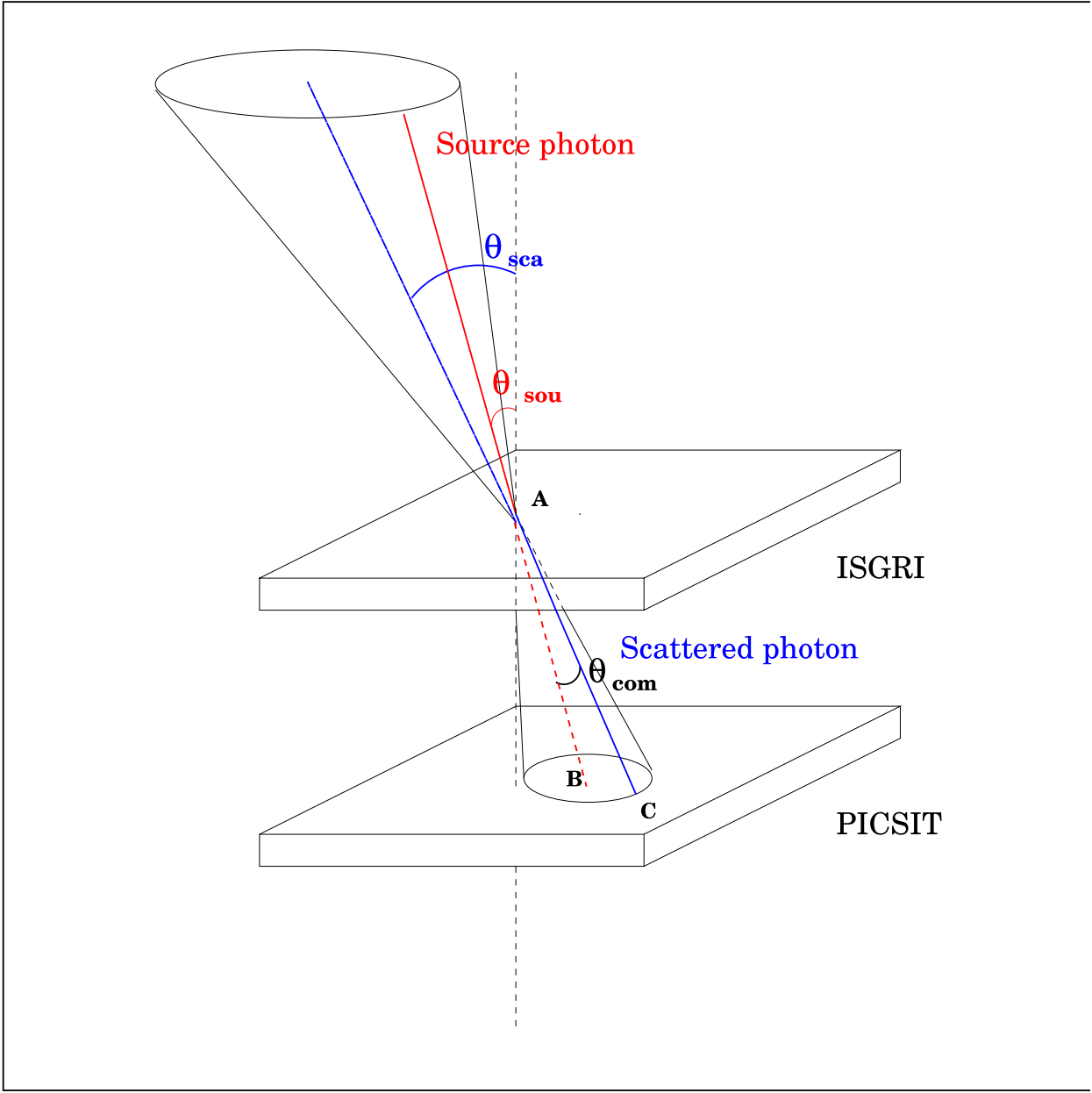}
\caption{Forward scattering of a photon in the IBIS/Compton mode.
An incident photon in red is scattered in ISGRI and absorbed in
PICsIT (in blue). The energy and position measurements in A and C
allow the derivation of the two angles: $\theta_{com}$ and
$\theta_{sca}$.\label{CompSca}}
\end{figure}
\clearpage
\section{Principle of a Coded Aperture Compton Telescope}

In a Compton telescope, consisting of two detector layers,
gamma-ray photons are Compton scattered in one detector and absorbed
in the second. The locations and energy deposits of each interaction
are measured, as illustrated in Figure \ref{CompSca} for the IBIS
detectors. The direction of the scattered photon, $\overrightarrow{u}_{sca}$,
is determined from the interaction locations
in the two detectors. The Compton scattering angle, $\theta_{com}$,
is measured from the energy deposits, $E_{1}$ and $E_{2}$, recorded
in the two detectors and given by, for a forward scattering :
\begin{equation}
\cos \theta_{com} = 1-\frac{m_ec^2}{E_{2}} +
\frac{m_ec^2}{E_{1}+E_{2}}
\end{equation}
where $m_{e}c^{2}$ is the electron rest mass energy. In a
Compton telescope, the direction of the incoming gamma-ray photon
lies on the edge of a cone, the Compton cone, with axis $\overrightarrow{u}_{sca}$ and aperture
$\theta_{com}$. The density distribution of all the projected
event circles, intersection of the cone with the celestial sphere, 
allows to reconstruct sky maps and to locate
sources. Source polarization can also be measured
since the scattering azimuth is related to the polarization
direction.  The angular resolution of the telescope depends on the
accuracy of the $\theta_{com}$ and $\overrightarrow{u}_{sca}$ determinations
and thus depends on the energy resolution and pixel sizes in each
of the two detectors. Furthermore, background is hard to subtract
and, as in optical cameras, several effects distort images. Using
a coded mask to reconstruct sky images effectively addresses most
of these difficulties.

In coded aperture telescopes, the source radiation is spatially
modulated by a mask of opaque and transparent elements. The
projection of the mask shadow recorded with a position sensitive
detector produces a shadowgram. This allows simultaneous
measurement of source plus background flux (shadowgram area
corresponding to the mask holes), and background flux (shadowgram
area corresponding to the opaque elements) \citep{Caroli1987}. The
background is removed in the deconvolution of the shadowgram using
the mask pattern. Mask patterns are designed to allow each source
in the field of view to cast a unique shadow on the detector in
order to avoid ambiguities in the reconstruction of the sky image.
The energy-independent angular resolution is given, in the thin mask limit
 (see the discussion for thicker masks above), by $d\alpha =
arctan(\frac{C}{H})$, where $C$ is the size of a mask element and
$H$ is the distance between the coded mask and the detector. In
such a telescope the field of view is limited and depends on the
mask dimension, the detector dimension and the mask-detector
distance. The total field of view is divided in two parts:

\begin{itemize}
\item FCFOV (Fully Coded Field Of View) for which the source
radiation is completely modulated by the mask. This field exists
only when the mask is larger than the detector.
\item PCFOV
(Partially Coded Field Of View) for which only a fraction of the
source radiation is modulated by the mask.
\end{itemize}

The principle of a CACT takes advantage of both techniques. It is
composed of a coded mask and two detector planes between which
photons are Compton scattered. The flux on one detector is
spatially modulated by the mask pattern. The sky image is obtained
by a simple deconvolution of this shadowgram. Compton events that
are incompatible with a given source direction can be discarded
from the shadowgram, so CACT can be regarded as a coded mask
telescope where the Compton kinematics are used to reduce the
background.

Then, two cases are possible: Either ones want to study a given source with a known position $\overrightarrow{u}_{sou}$ 
in the celestial sphere or ones wish to make an image of a given field of view.
In the first case, we can select, using Compton kinetics, events which fulfill the following condition:

\begin{equation}
\overrightarrow{u}_{sou}.\overrightarrow{u}_{sca} = cos(\theta_{com})
\end{equation}

within instrumental incertainties, as, by definition, $\theta_{com}$ is the angle between the source and the scattered directions. In the case of a isotropical background, Eq. 2 enable typically to remove more than $90\%$ of the  Compton forward background events, while keeping $90\%$ of the Compton forward source events, in the 200 keV - 1 MeV energy range.

When we want to study sources over a given field of view, the more conservative way of removing background events using the Compton kinetics, is to remove all events whose Compton cones, within uncertainties, do not intersect the field of view. This condition can be readily written in the plane containing the telescope axis and the source direction (for a forward scattering). Indeed, if we consider the case of a conical field of view of semi-angle $\theta_{FOV}$, the selection condition means that the angle between the source direction and the telescope axis $\overrightarrow{u}_{tel}$, called $\theta_{sou}$, should be greater than the angle $\theta_{FOV}$. $\theta_{sou}$ can be easily computed from the Compton angle and the scattered photon ones ($\theta_{sca}$ such as $cos(\theta_{sca}) = \overrightarrow{u}_{sca}.\overrightarrow{u}_{tel}$). The background rejection condition become then:

\begin{equation}
\theta_{sou} = (\theta_{sca}-\theta_{com}) \geq \theta_{FOV}
\end{equation}

A similar formula can be obtained in the backward scattering case.

\section{The INTEGRAL IBIS/Compton mode}

\subsection{The IBIS telescope as a CACT}

The IBIS instrument \citep{Ubertini2003} is one of the two major
coded aperture telescopes on board the ESA INTEGRAL gamma-ray
observatory launched on October $17$ $2002$. It consists of a dual
detection layer designed and optimized to operate in the energy
range between $\sim15~\hbox{keV}$ and $10~\hbox{MeV}$. The upper
detector layer, ISGRI, covering the energy range from $\sim
15~\hbox{keV}$ to $1~\hbox{MeV}$, is made of  $128\times128$
Cadmium--Telluride (CdTe) semiconductor detectors
\citep{Lebrun2003}. The lower detector layer, PICsIT, operating in
the energy interval from $\sim 190~\hbox{keV}$ to $10~\hbox{MeV}$,
is made of $64\times64$ Cesium--Iodide (CsI) scintillating
crystals \citep{Labanti2003}. Events from these two detection
layers are time stamped and an on board Hardware Event Processing
Unit (HEPI) can associate the ISGRI and PICsIT events if their
arrival times differ by less than a given time coincidence window 
(actually 3.8 $\mu s$). In the following,
these events are referred to as tagged Compton events. The detector
spectral drifts (gain changes) can be monitored with a $^{22}Na$
On Board Calibration Unit (OBCU). The detector layers are actively
shielded, encased on all but the sky side by bismuth germanate
(BGO) scintillator elements. It is also passively shielded from
the low energy celestial background with tungsten and lead foils.
The coded mask is made of 16 mm thick tungsten elements of 11.2 mm
by side. This thickness guarantees a 50\% modulation at 1 MeV.
Placed $3.2~\hbox{m}$ above the CdTe detector plane, this mask
ensures a 12 arcmin angular resolution over a $29°\times29°$
PCFOV. Composed of two detector planes (ISGRI and PICSIT) able to
work in coincidence and covered with a coded mask, the IBIS
telescope is the first in flight CACT.

\subsubsection{Event types} \label{Evtype}

Tagged Compton events from the celestial source under study can be of two kinds:

\begin{itemize}
\item true Compton events,
\item or spurious events, where two independent ISGRI and PICsIT events, one of them coming from the source, fall by chance in the Compton coincidence time window, and are recorded falsely as a true Compton event.
\end{itemize}
Below 500 keV, the vast majority of Compton scatters corresponds
to forward scattered events (ISGRI $\rightarrow$ PICSIT). With
increasing energy, photons can pass through ISGRI without any
interaction, interact in PICSIT, and scatter back onto the ISGRI
detection layer. In some cases, more than one scattering occurs.
Multiple interactions in ISGRI are, however, discarded on board.
In this paper we will use only the events that underwent a forward
scattering in ISGRI with a single energy deposit in PICsIT.

\subsubsection{spectral resolution}

In standard Compton telescopes, the spectral resolutions of each
detector are key parameters since they directly affect the angular
resolution which, in turn, governs the sensitivity. For CACT, the
angular resolution is driven by the mask geometrical properties (C
and H), but the sensitivity strongly depends on background
rejection. The latter is based on measuring $\overrightarrow{u}_{sca}$ and
$\theta_{com}$. The uncertainty on $\theta_{com}$, $\delta\theta_{com}$

\begin{equation}
\delta\theta_{com} = \frac{m_ec^2}{E^2sin\theta} +
\sqrt{\delta E_{1}^2+ \left(\frac{E_1^2}{E_2^2} + 2\frac{E_1}{E_2}\right)^2 \delta E_{2}^2}
\end{equation}

where $\delta E_{1}$ and $\delta E_{2}$ are the energy resolution of the first and second detector layers, respectively, is larger in IBIS than that on
$\overrightarrow{u}_{sca}$ which relates to pixel size.
We have used the Compton data tagged as
calibration events by the On Board Calibration Unit to measure the on board spectral resolution of the IBIS Compton mode. 
The FWHM of the two lines of the
$^{22}$Na source (511 keV and 1274 keV) and the resulting energy
resolution are presented in Table \ref{Table1}.
\clearpage
\begin{table}[!h]
\caption{In flight IBIS Compton mode energy resolution}

\label{Table1}
\[
    \begin{array}{cc}
    \hline
\noalign{\smallskip}

\rm Energy (keV) & \rm Energy~  resolution~ (\%  ~FWHM) \\
\hline
\noalign{\smallskip}
\rm  511 &  20      \\
\rm 1274 &  15      \\
\hline
\end{array}
   \]
\end{table}
\clearpage
\section{Imaging the sky with the IBIS Coded Aperture Compton Telescope}

In this section we focus on imaging analysis and performance of
the IBIS/Compton mode.

\subsection{The IBIS Compton mode imaging analysis}

\subsubsection{Events selection}

The first step to analyze the IBIS Compton mode data is to apply
selections on the events: selection in energy
(generally between 200 keV and 1 MeV), and selection of ISGRI events in rise time
between 0.6 and 3.8 $\mu s$ (see Lebrun et al., 2003, for a description of ISGRI 
data).

Then, we remove background events using the Compton kinetics, as described 
in paragraph 2. As discussed there, depending on the purpose, there are two types of 
selection.
\begin{itemize}
\item {\em The field of view selection}: The IBIS field of view semi-angle being 
$\theta_{FOV} \simeq
15^{\circ}$, only photons with $\theta_{sca}-\theta_{com} < 15^{\circ}$ are kept.

\item {\em The dedicated source selection}: for a source of known
direction $\overrightarrow{u}_{sou}$, a more restrictive selection given by Eq. 2 
is applied. This condition can be rewritten as:
$|\overrightarrow{u}_{sou}.\overrightarrow{u}_{sca} - cos(\theta_{com})| < \delta_{lim}$ where $ \delta_{lim}$ is related to the instrumental error.
\end{itemize}

We compute values of $\delta_{lim}$ in order to maximize the source
signal to noise ratio, using ground calibration measures.
We have used the Compton
events obtained from three on-axis calibration sources, namely
$^{133}$Sn ($392$~keV), $^{22}$Na ($511$~keV), and $^{137}$Cs
($662$ keV). In fact, for a on-axis source, the telescope axis 
and source direction coincide, so 

\begin{equation}
\overrightarrow{u}_{sou} = \overrightarrow{u}_{tel} 
\end{equation}

then, from Equation 2, we get

\begin{equation}
cos(\theta_{com}) = \overrightarrow{u}_{sou}.\overrightarrow{u}_{sca} = \overrightarrow{u}_{tel}.\overrightarrow{u}_{sca} = cos(\theta_{sca})
\end{equation}

by definition of $\theta_{sca}$. Equation 2 then simplify to :

\begin{equation}
\Delta \theta = \theta_{com} -\theta_{sca}= 0
\end{equation}

\begin{figure}[!h]
\epsscale{1.2}
\plotone{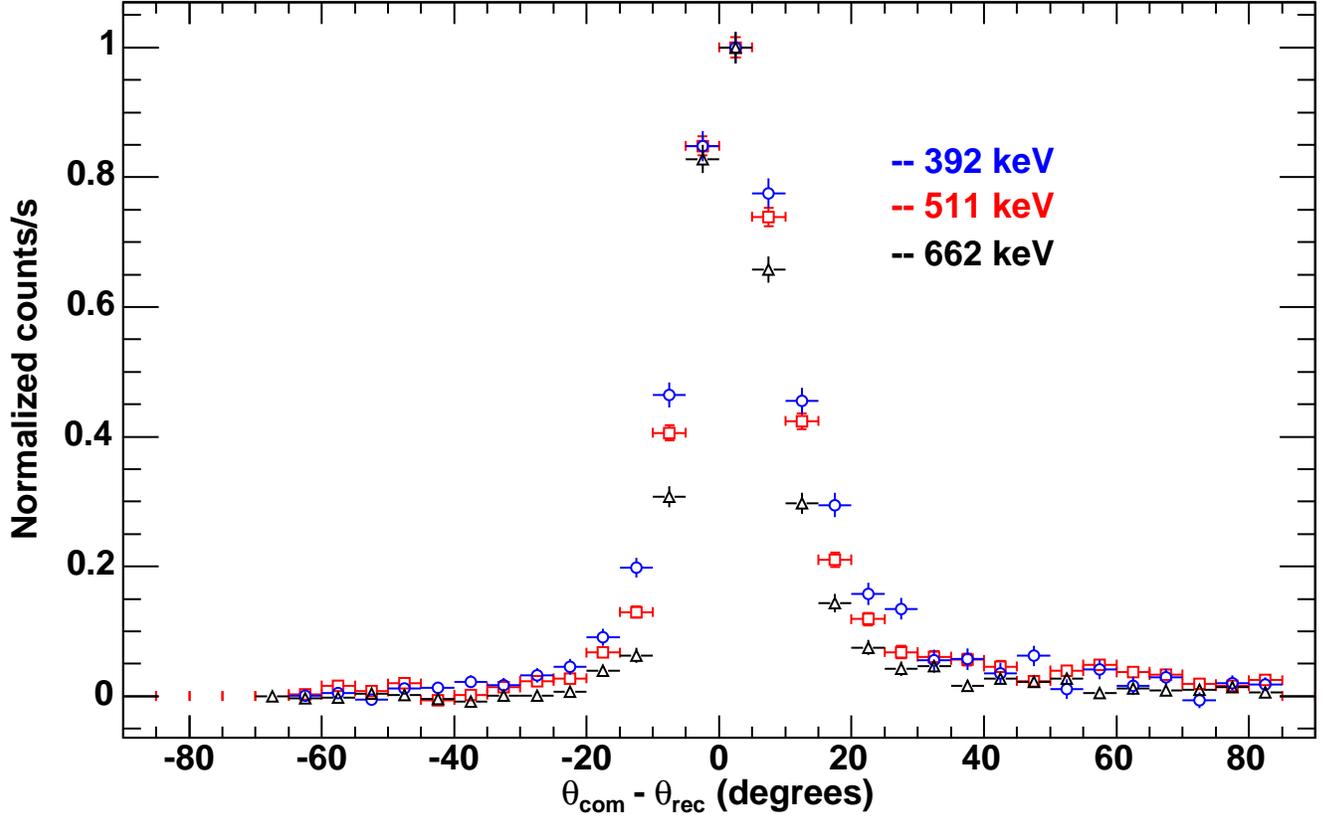}
\caption{Distribution of Compton events with the angular shift
$\Delta \theta = \theta_{com}-\theta_{sca}$, after spurious events
removal, using on-ground
calibration data from $^{133}$Sn ($392 keV$), $^{22}$Na ($511
keV$), $^{137}$Cs ($662 keV$) sources ($\theta_{rec}$ in this figure is equal to $\theta_{sca}$ in the text). \label{DeltaPhiCalAE}}
\end{figure}

Figure \ref{DeltaPhiCalAE} shows the angular shift $\Delta \theta$ diagrams. This distribution, centered on zero, is not a Dirac distribution because of instrumental incertainties. Also, this distribution narrows with energy due to a better
reconstruction of $\theta_{com}$, linked to a better Compton mode energy resolution at
high energy. Yet, this variation with energy is small and the optimal choice of $\delta_{lim}$ (related to the width of the distribution shown in figure 2) has been checked not to change much with energy between 200 keV and 1 MeV. 

\begin{figure}[!h]
\epsscale{1.2}
\plotone{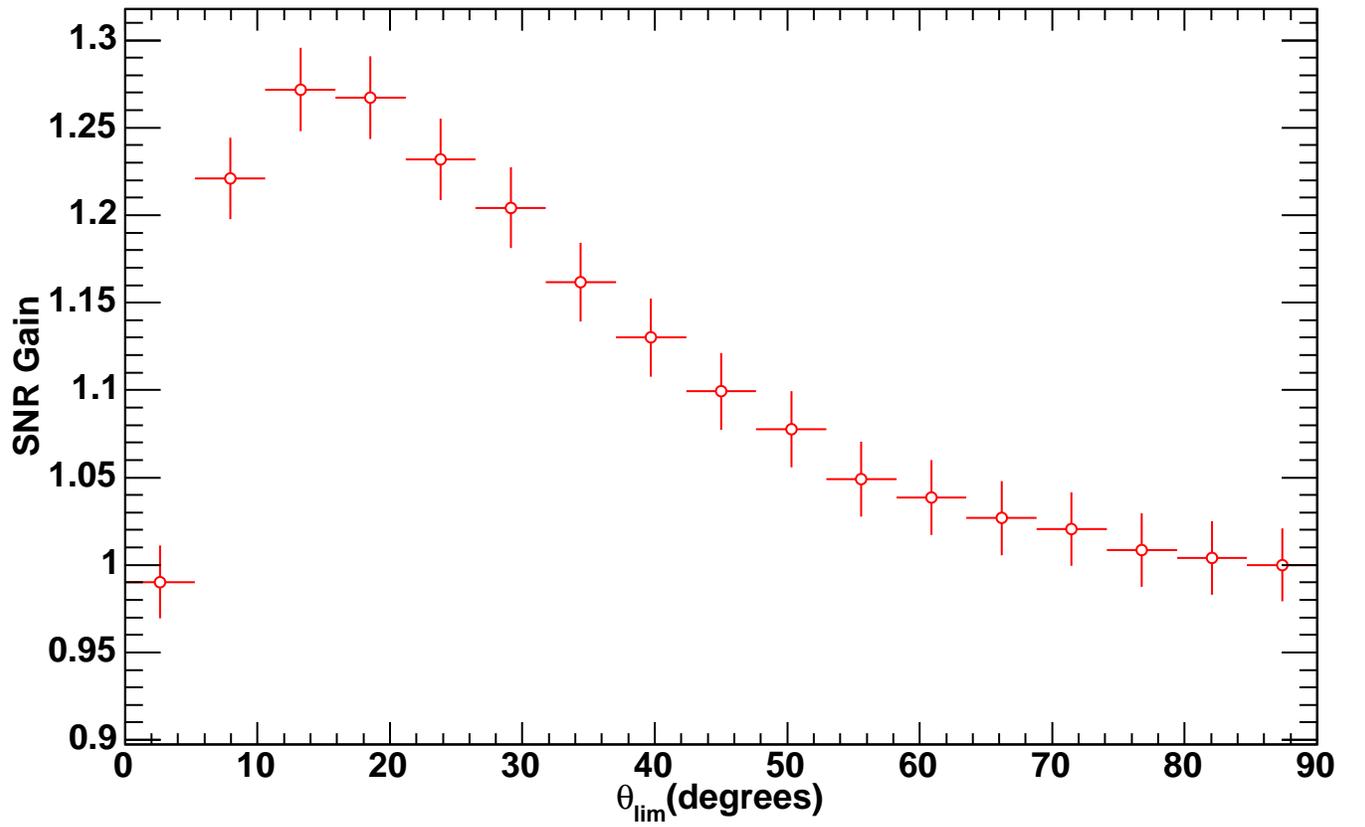}
\caption{Evolution of the signal to noise ratio gain with
$\theta_{lim}$ for the $^{133}$Sn calibration source at 392 keV on
axis.\label{CompSelGain}}
\end{figure}

Figure \ref{CompSelGain} illustrates how the signal to noise ratio varies with the allowed range of 
$\Delta \theta \in [-\theta_{lim},\theta_{lim}]$ for the $^{133}$Sn calibration source. The best value of $\theta_{lim}$ at $392$~keV is around $10-12^{\circ}$.

\clearpage

\clearpage

\subsubsection{Spurious events subtraction}

Spurious events are generally neglected in standard Compton
telescopes. Their uniform distribution does not induce false
source detection. But the situation for CACT is different:

Indeed, spurious events are composed by random events on the two layers.
For a bright source, like Crab, the source low energy photons have a 
important contribution in the detectors count rates. So, the probability
that one of these random events is in fact a detected low energy 
photon from the source is quite high. As this photon has the mask signature
 induced by the source, it is not subtracted during the deconvolution 
process, and wrongly participate to the source flux determination.
So, we have to take into account the spurious events contribution with high 
accuracy in order to get a correct estimate of the source flux. 

The dedicated source selection has the strong advantage of 
largely reducing the number of spurious events, as most of
them does not obey Equation 2, but their remaining contribution 
is not negligible. A statistical method must be applied to
evaluate and subtract them.

To do so, we make use of the ISGRI and PICSIT single events recorded in
the same observation, that is possibly having the source signature, to
artificially associate them and create a sample of spurious
events, called hereafter the ``fake spurious events sample''. 
We apply to this sample the same selections in energy, rise time, and
scattering angle as described above in order to produce a fake
shadowgram. The latter is scaled, by a scaling factor called hereafter
``spurious factor'' or ``$\alpha$'', to the number of spurious
events recorded during the observation and then subtracted from the 
Compton data shadowgram. 

The spurious count rate, $N_{Sp}$ scales with the
width of the time coincidence window ($\sim 2 \Delta T_{e}$), the total
number of PICSIT events ($N_{PICSIT} = N_P + N_{OBCU}$ from PICSIT
simple and multiple detections, and from calibration events), and
the total number of ISGRI events ($N_{ISGRI} = N_I +N_{Sp}$ from ISGRI
single events and spurious events). The calibration
 events in ISGRI are tagged and discarded on board. Using 
Poisson statistics in the coincidence window, one obtains the number of 
spurious events:

\begin{equation}
N_{Sp} =(1 - e^{-(2\Delta T_e -\delta T)(N_{PICSIT})})N_{ISGRI}
\end{equation}
where $\delta T$ is the on-board time resolution, of the order of 250 ns. Yet, one
measures only $N_I$, so the scaling factor is
\begin{equation}
N_{Sp}/N_I = (e^{(2\Delta T_e -\delta T)(N_{PICSIT})}-1)
\end{equation}

One has to further correct this factor for the multiple PICSIT
events in order to get the scaling factor for the single
spurious events only. For a proportion of PICSIT single events
of

\begin{equation}
\beta =
\frac{N_{P}^{simple}+N_{OBCU}}{N_{P}^{simple}+N_{P}^{multiple}+N_{OBCU}}
\end{equation}

($\beta$ is of the order of $80 \%$ between 200 keV and 1 MeV for IBIS), one gets the spurious factor

\begin{equation}
\alpha = \beta (e^{(2\Delta T_e -\delta T)(N_{PICSIT})}-1)
\end{equation}

\begin{figure}[!h]
\epsscale{1.2}
\plotone{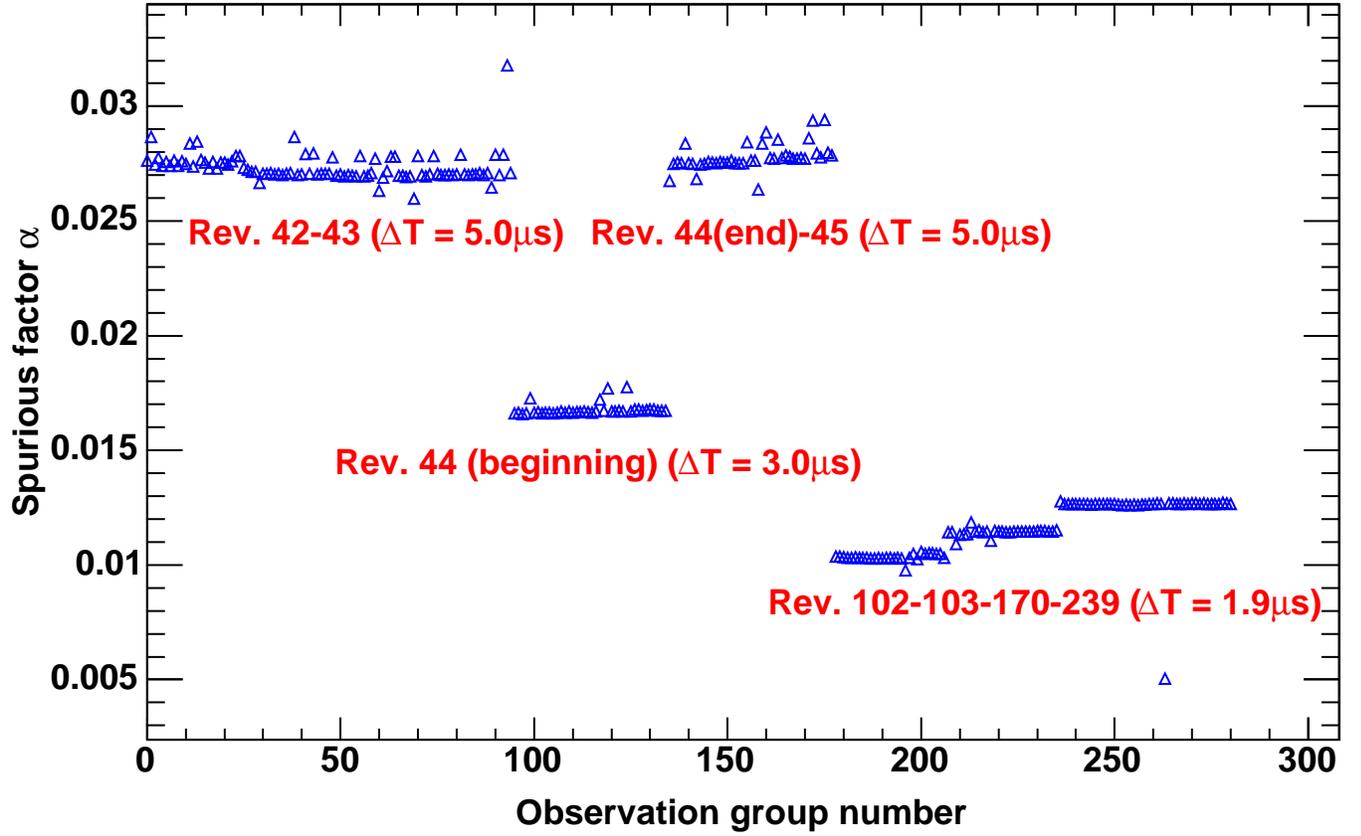}
\caption{Evolution of the spurious factor for different values of
the time coincidence window as derived from Crab pulsar
observations.\label{SpurFact}}
\end{figure}

Figure \ref{SpurFact} shows the evolution of the spurious factor
for Crab observations through the mission lifetime. It varies from
$2.82~ \%$ early in the mission (revolution $39$) when $\Delta T$
was about $5.0 ~\mu s$, to around $1.1~ \%$
after revolution $102$ when $\Delta T_e$ was decreased to $1.9 ~\mu
s$. $\alpha$ is quite constant for a given coincidence window
during a short time, but rises on longer periods (Rev.
$102-103-170-239$), due to the increase of the background flux,
following the solar cycle.
\clearpage
\clearpage
The next step is to correct the resulting shadowgram (after
subtraction of the spurious events) for its non-uniformity and to
deconvolve it.

\subsubsection{Uniformity correction}

Compton mode shadowgrams are not spatially flat. The count rate
 falls near the edges because we lose the events which scatter
at the edge of ISGRI and miss PICSIT. This
non-uniformity is magnified by the decoding process so, if not corrected,
 strong systematic structures may result in the deconvolved images with
spatial frequencies similar to those originally present in the
shadogram. The shadowgram, D, consists of a source
component, with count rate S and a spatial response map $R_S$, and 
a background component, with count rate B and response map $R_B$, thus

\begin{equation}
D = S \times R_{S} + B \times R_{B} 
\end{equation}

The modulation from the mask pattern is weak compared to the
larger scale deformations we study here. To compare $R_{S}$ and
$R_{B}$ in the same conditions, we need in-flight data from a
source strong enough as well as background. Whereas we have used the
in-flight background distribution, there is no such source 
observed in-flight in Compton mode, so we used data from on-ground calibration 
to determine $R_{S}$.
\clearpage
\begin{figure}[!h]
\epsscale{1.2} \plotone{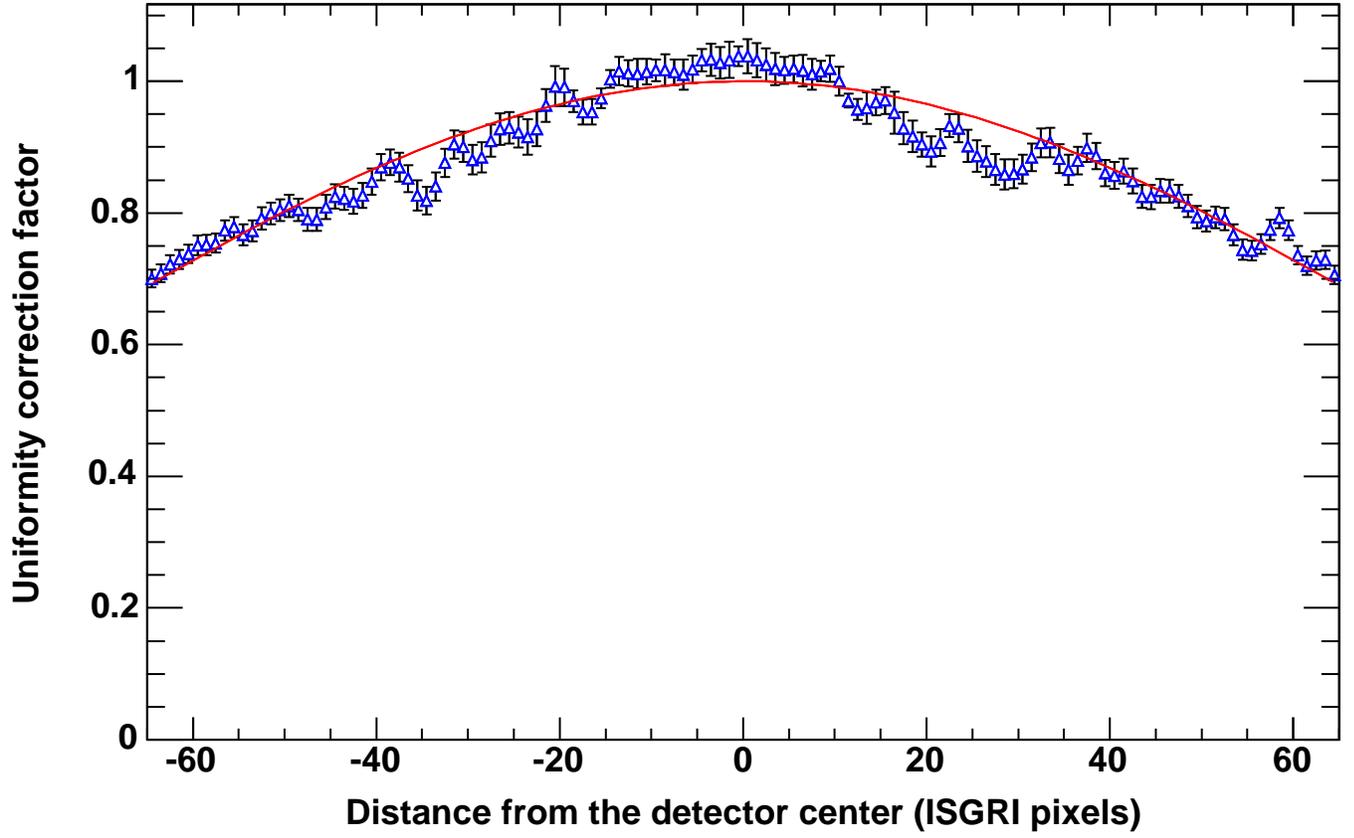}
\caption{Distribution of background events with distance from the
detector center, folded over the azimuthal direction. This distribution is well fitted by a 2D gaussian
shown in red.\label{unif}}
\end{figure}
\clearpage
Both the background and source response maps are well fitted by a
2D gaussian function (see Figure \ref{unif}). The results are
presented in Table \ref{Table3} for the in-flight background
measurements and the on-ground calibration source data for on-axis
sources. The background and source response maps in several energy
bands are found to be consistent. Off-axis sources have also
been studied and display equivalent response maps.
\clearpage
\begin{table}
\caption{response maps gaussian fits}

\label{Table3}
\[
    \begin{array}{ccc}
    \hline
\noalign{\smallskip}

\rm Source~ & \rm Energy (keV) & \rm \sigma (pixels) \\
\hline
\noalign{\smallskip}
\rm  ^{113}Sn & 392 &  73 \pm 3      \\
\rm  ^{22}Na & 511 &  74 \pm 3      \\
\rm  ^{137}Cs & 662 &  71 \pm 4      \\
\rm  ^{54}Mn & 835 &  72 \pm 2      \\
\rm  ^{54}Zn & 1120 &  73 \pm 3      \\
\rm  Background & 200-5000 &  74 \pm 3      \\
\hline
\end{array}
   \]
\end{table}
\clearpage
Residual background deformations on
smaller scales, similar what is observed in ISGRI images \citep{Terrier2003},
are still present; their correction is under study.

The final step is therefore to deconvolve the corrected shadowgram
$D/R$, renormalized to the total number of events, to
reconstruct the source flux, using standard deconvolution techniques.

\subsubsection{Image deconvolution}

Representing the mask with an array M of 1 (transparent) and 0
(opaque) element, the detector plane by an array D,  and denoting by
$G^{+}$ and $G^{-}$ the decoding arrays related to the coded mask (see \citep{Goldwurm2003}), the image
deconvolution in the fully coded field of view (FCFOV) can be
extended to the total (fully coded and partially coded) by
performing the correlation of D in a non cyclic form with the G
array extended and padded with 0 elements outside the mask \citep{Gros2003}. Since the number of correlated (transparent and opaque) elements in the
partially coded field of view is not constant as in the FCFOV, the
sum and subtractions for each sky position must be balanced and
renormalized. The sky flux map is given by :

\begin{equation}
F = \frac{(WD*G^{+})-(WD*G^{-})\frac{(W*G^{+})}{(W*G^{-})}}{(W*G^{+}M)-(W*G^{-}M)\frac{(W*G^{+})}{(W*G^{-})}}
\end{equation}

and the variance map by:

\begin{equation}
V = \frac{(W^{2}D*G^{+2})+(W^{2}D*G^{-2})(\frac{(W*G^{+})}{(W*G^{-})})^{2}}{((W*G^{+}M)-(W*G^{-}M)\frac{(W*G^{+})}{(W*G^{-})})^{2}}
\end{equation}

In the previous formulae, the W matrix removes dead or noisy pixels.

\begin{figure}[!h]
\epsscale{1}
\plotone{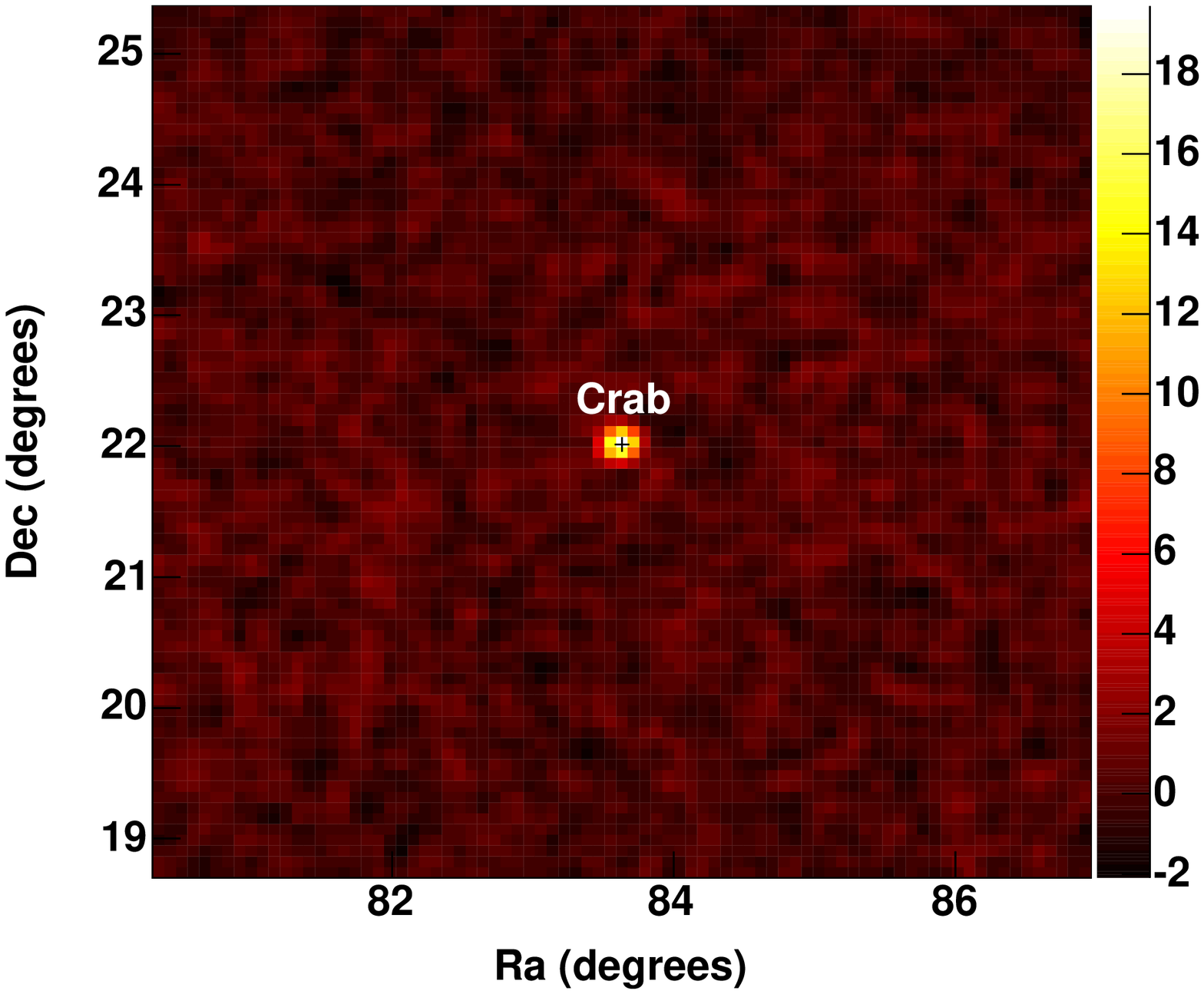}
\caption{Deconvolved significance map for the Crab pulsar using
the Compton mode between $200~keV$ and $500~keV$ for an exposure
time of $300~ks$.\label{CompCrabIma}}
\end{figure}

In the FCFOV, the variance is quite uniform and equals the total
number of detector counts. All this analysis procedure can be
easily carried out by means of Fast Fourier transforms. The result
on a Crab pulsar observation is presented on Figure
\ref{CompCrabIma}.

\begin{figure}[!h]
\epsscale{0.5}
 \centering 
 \leavevmode 
 \includegraphics[width=200pt,angle=270]{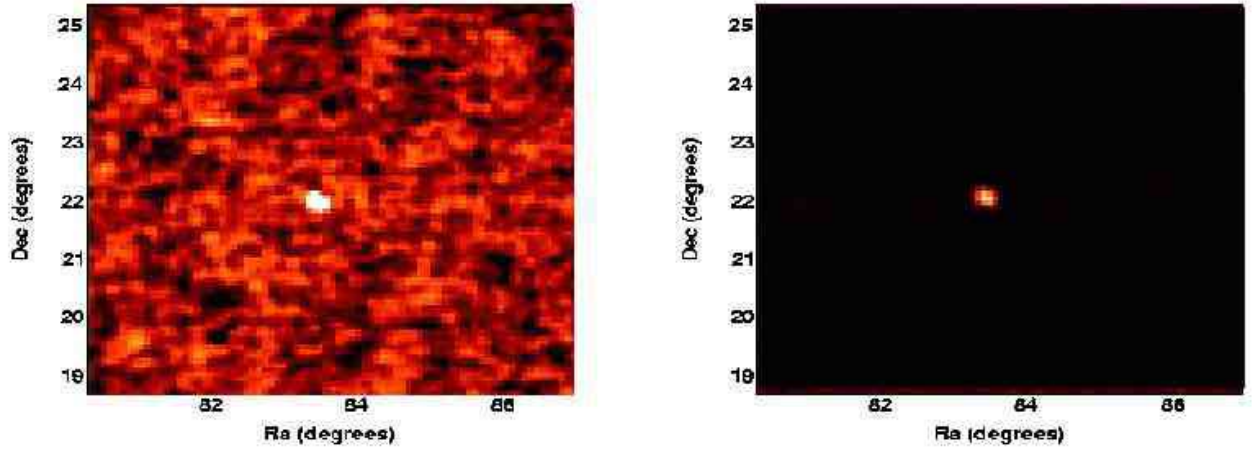}%
\caption{left: ISGRI significance map of the Crab pulsar  between $200~keV$ and $500~keV$. right: Significance map computed from the fake spurious events sample, and computed with the same algorithms as used for the map shown in Figure \ref{CompCrabIma}. \label{SpurCrabIma}}
\end{figure}

On Figure \ref{SpurCrabIma}, on the left, we show for comparison an ISGRI alone significance map, in the same energy range. As expected, due to its low thickness (2 mm), ISGRI is much less sensitive that the Compton mode between 200~keV and 500~keV, showing that the Compton mode is a real extension at high energy of the ISGRI capabilities. On Figure \ref{SpurCrabIma}, on the right, we show the significance map made with the fake spurious events sample, using the same algorithms as the ones used for Figure \ref{CompCrabIma}. It is clear there that this map is dominated by low energy events coming from the Crab, as described in part 4.1.2.

\clearpage
\clearpage
\clearpage
\clearpage

\subsubsection{Angular shift diagrams as a check for the analysis method}

\begin{figure}[!h]
\epsscale{1.2}
\plotone{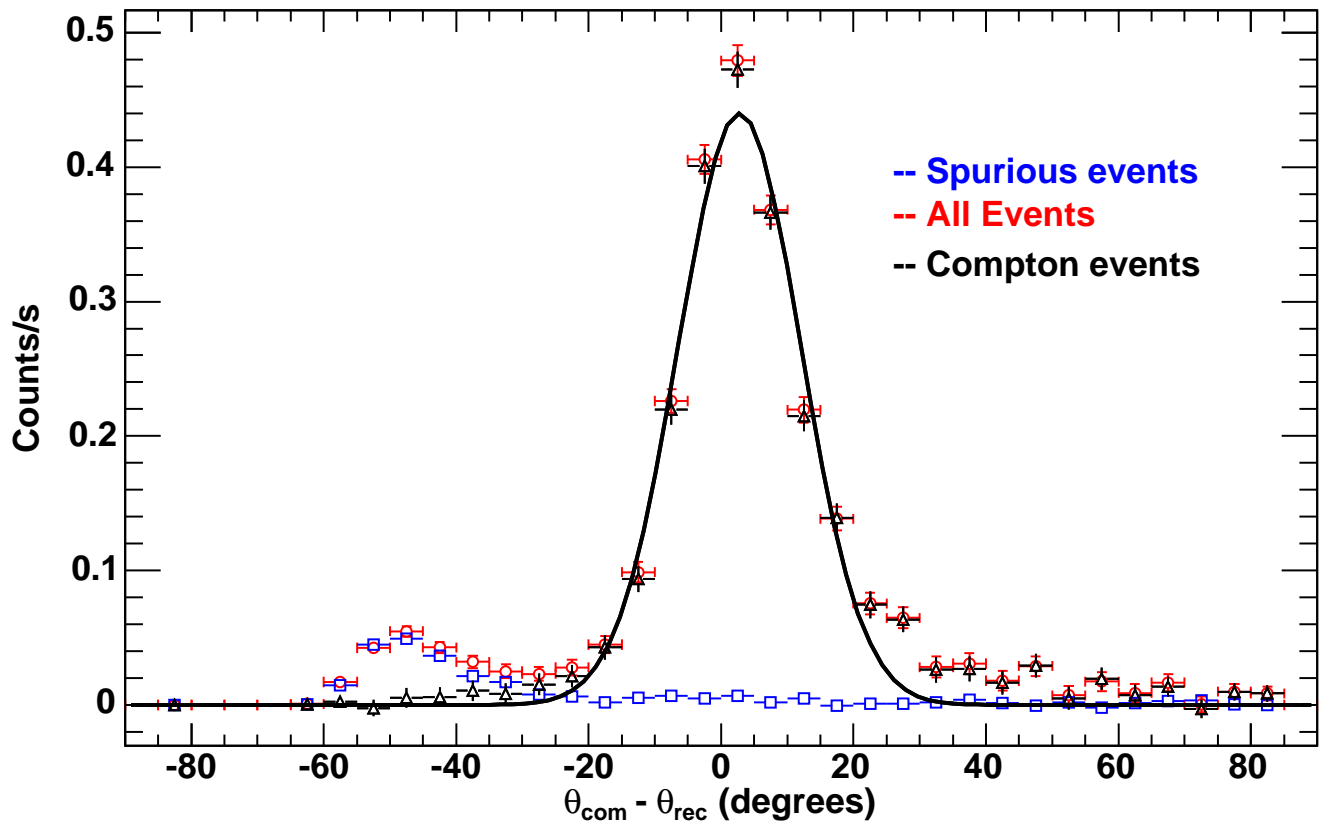}
\caption{Angular shift distribution of events for a calibration
source of $^{133}$Sn at 392 keV during on ground
calibration. Red data points represent all the Compton data (real Compton plus spurious data). Blue data points show the spurious contribution, which peaks at negative offset, and black data points are the derived Compton ones. The line is a gaussian fit ($FWHM \sim 19^{\circ}$) to this derived Compton data.\label{DeltaPhiCalib}}
\end{figure}

Angular shift diagrams illustrate the effectiveness of the
spurious event subtraction. One can use Compton tagged events from
an on-axis calibration source and analyze them in regularly spaced $\Delta
\theta$ bins. Then we select them in energy and rise time as above. 
Their shadowgram are corrected for the spatial response and
deconvolved to get the total source count rate displayed in red in
Figure \ref{DeltaPhiCalib}. The corresponding constructed  spurious event sample has
been analyzed in the very same way and its count rate per $\Delta
\theta$ bin, scaled by the measured $\alpha$ factor, is displayed
in blue, showing that the spurious factor is adequate. The angular
shift $\Delta \theta$ distribution of real Compton events
(after subtraction of the spurious ones) is well centered around
zero and fall to zero for $|\Delta \theta| \ge
19^{\circ}$ whereas the spurious distribution is clearly offset to
negative values, as expected because most spurious events have a
low energy deposit in ISGRI.

\begin{figure}[!h]
\epsscale{1.2}
\plotone{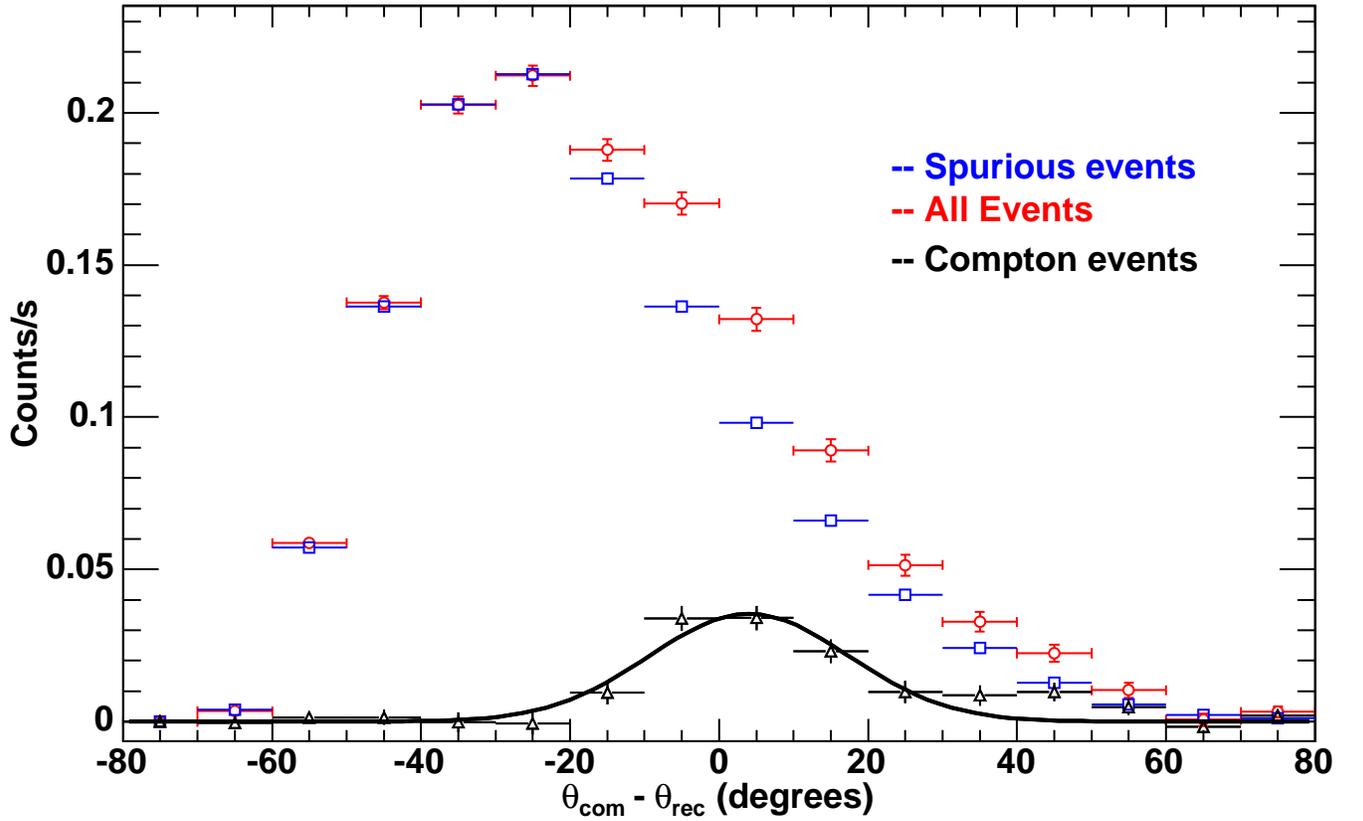}
\caption{Crab count rate between $200$ keV and $500$ keV in
different $\Delta \theta$ bins. The total observation time is
about $700$ ks. Red data points represent all the Compton data (real Compton plus spurious data). Blue data points show the spurious contribution, and black data points are the derived Compton ones. The line is a gaussian fit to the Compton data.\label{DPhi300500}}
\end{figure}

For celestial $\gamma$-ray sources above 200 keV, the spurious
rate dominates over the source rate. Several Crab on-axis observations have
been used to construct a $\Delta \theta$ diagram, using the
variance weighted sum of the flux at the source position in each
sky image. Thanks to the coded mask background subtraction
capabilities, only the true Compton and spurious contributions as defined in section \ref{Evtype}, are
visible on Figure \ref{DPhi300500}. The spurious component
severely dominates,  its negative offset being well marked. True 
Compton events are around zero, as foreseen, and the small flux excess
of events
for $\Delta \theta \sim 20^{\circ}-40^{\circ}$ is due to backward
scattered events.
\clearpage

\clearpage
\subsection{The IBIS Compton mode sensitivity}

\begin{figure}[!h]
\epsscale{1.2}
\plotone{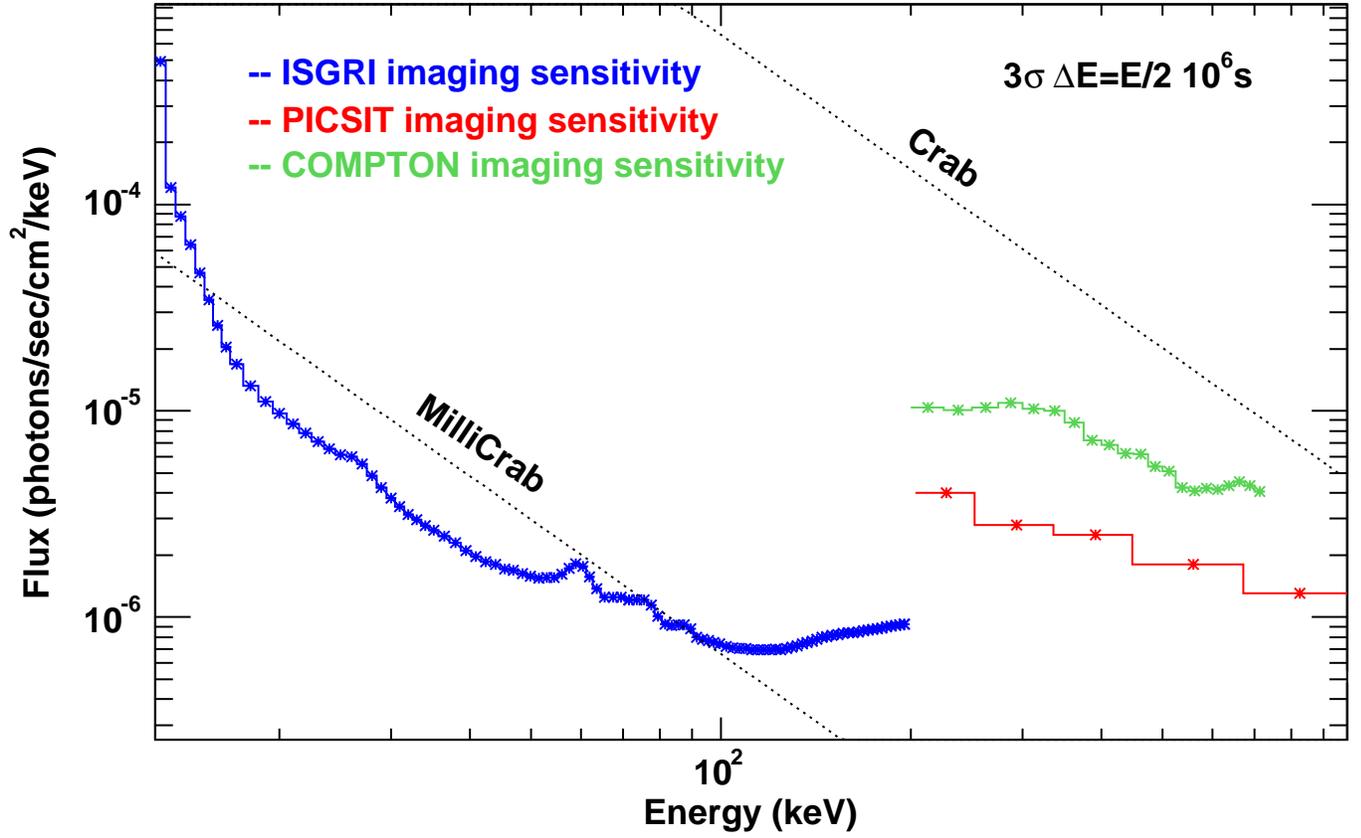}
\caption{Sensitivity of the IBIS Compton mode compared to the ISGRI
and PICSIT sensitivity.\label{Sensi}}
\end{figure}

The analysis method described above has been applied to evaluate the signal to
noise ratio of the Crab pulsar in different energy bands. The
sensitivity of the IBIS Compton mode is presented on Figure
\ref{Sensi}. It is greater than that of PICSIT for a
similar angular resolution. 
Yet, unlike PICSIT, the Compton mode
has no major strong background problems, allows to study photons up 
to a few MeV in very small energy bands, in particular
around spectral lines, with an angular resolution better than
 that of SPI. It also allows to perform polarization
studies and imaging studies of compact objects with a good timing 
resolution ($\sim 100 \mu s$).
\clearpage
\clearpage
The next step we foresee in our analysis is to incorporate
backward scattered events, PICSIT multiple events, and compute
background Compton correction maps (first order background
shadowgram from empty field observations and second order summed
sky images after source subtraction) to reduce the residual
structures in the response maps.

An important goal of the IBIS Compton mode is also polarimetry. The
interest of the astrophysics community to such detection is
growing. It is in fact a powerful and a direct tool to constrain
theoretical model on GRB, pulsars, solar flares, etc. 

The calibration and results of the IBIS Compton mode polarimeter will 
be presented in a forecoming paper.

\section{Conclusions}

The IBIS Compton mode is functional and provides a new efficient
means to observe the sky at energies beyond $\sim 190~\hbox{keV}$ 
up to a few MeV. With only forward scattered events and thus thanks 
to the ISGRI shadowgram, we can reconstruct images with high spatial resolution
taking advantage of the coded mask aperture system. We have
devised a scheme for subtracting the large contribution from
spurious coincidences between the two detector planes. The
resulting sensitivity, evaluated with in-flight data from the Crab
pulsar, opens new perspectives for polarimetric and imaging
studies in the 0.2-5 MeV energy band.


We thank the anonymous referee for interesting questions and helpful suggestions that significantly improved the whole paper. 


\end{document}